\documentclass[10pt]{bmc_article}    

\usepackage{cite} 
\usepackage{hyperref}  
\usepackage{ifthen}  
\usepackage{multicol}   
\usepackage[utf8]{inputenc} 
\usepackage{graphicx}

\newboolean{publ}

\newenvironment{bmcformat}{\fussy\setboolean{publ}{true}}{\fussy}

\begin{document}
\begin{bmcformat}

\title{Analyzing huge pathology images with open source software}


\author{Christophe Deroulers\correspondingauthor$^1$%
       \email{Christophe Deroulers\correspondingauthor - deroulers@imnc.in2p3.fr}%
      \and
         David Ameisen$^2$%
         \email{David Ameisen - david.ameisen@gmail.com}
      \and
         Mathilde Badoual$^1$%
         \email{Mathilde Badoual - badoual@imnc.in2p3.fr}
      \and
         Chlo\'e Gerin$^{3,4}$%
         \email{Chlo\'e Gerin - chloe.gerin@u-psud.fr}
      \and
         Alexandre Granier$^5$%
         \email{Alexandre Granier - alexandre.granier@mri.cnrs.fr}
       and 
         Marc Lartaud$^{5,6}$%
         \email{Marc Lartaud - lartaud@cirad.fr}%
      }
      

\address{%
    \iid(1)Univ Paris Diderot, Laboratoire IMNC, UMR 8165 CNRS, %
        Univ Paris-Sud, F-91405 Orsay, France\\
    \iid(2)Univ Paris Diderot, Laboratoire de pathologie, %
        H\^opital Saint-Louis APHP, INSERM UMR-S 728, F-75010 Paris, %
        France\\
    \iid(3)on leave from CNRS, UMR 8165, Laboratoire IMNC, Univ %
        Paris-Sud, Univ Paris Diderot, F-91405 Orsay, France\\
    \iid(4)now at CNRS, UMR 8148, Laboratoire IDES, Univ Paris-Sud, %
        F-91405 Orsay, France and CNRS, UMR 8608, IPN, Univ Paris-Sud, %
        F-91405 Orsay, France\\
    \iid(5)MRI-Montpellier RIO Imaging, CRBM, F-34293 Montpellier, %
        France\\
    \iid(6)CIRAD, F-34398 Montpellier CEDEX 5, France
}%

\maketitle


\begin{abstract}
        \paragraph*{Background:} Digital pathology images are 
increasingly used both for diagnosis and research, because slide 
scanners are nowadays broadly available and because the quantitative 
study of these images yields new insights in systems biology. However, 
such virtual slides build up a technical challenge since the images 
occupy often several gigabytes and cannot be fully opened in a 
computer's memory. Moreover, there is no standard format. Therefore, 
most common open source tools such as ImageJ fail at treating them, and 
the others require expensive hardware while still being prohibitively 
slow.
      
        \paragraph*{Results:} We have developed several cross-platform 
open source software tools to overcome these limitations. The NDPITools 
provide a way to transform microscopy images initially in the loosely 
supported NDPI format into one or several standard TIFF files, and to 
create mosaics (division of huge images into small ones, with or without 
overlap) in various TIFF and JPEG formats. They can be driven through 
ImageJ plugins. The LargeTIFFTools achieve similar functionality for 
huge TIFF images which do not fit into RAM. We test the performance of 
these tools on several digital slides and compare them, when applicable, 
to standard software. A statistical study of the cells in a tissue 
sample from an oligodendroglioma was performed on an average laptop 
computer to demonstrate the efficiency of the tools.

        \paragraph*{Conclusions:} Our open source software enables 
dealing with huge images with standard software on average computers. 
Our tools are cross-platform, independent of proprietary libraries, and 
very modular, allowing them to be used in other open source projects. 
They have excellent performance in terms of execution speed and RAM 
requirements. They open promising perspectives both to the clinician who 
wants to study a single slide and to the research team or data centre 
who do image analysis of many slides on a computer cluster.

\end{abstract}

\ifthenelse{\boolean{publ}}{\begin{multicols}{2}}{}


\section*{Background}

Virtual microscopy has become routinely used over the last few years for 
the transmission of pathology images (the so-called virtual slides), for 
both telepathology and teaching~\cite{virtual-microscopy, 
stack-or-trash}. In more and more hospitals, virtual slides are even 
attached to the patient's file~\cite{openaccess-4, these-david-ameisen}. 
They have also a great potential for research, especially in the context 
of multidisciplinary projects involving e.g.\ mathematicians and 
clinicians who do not work at the same location. Quantitative histology 
is a promising new field, involving computer-based morphometry or 
statistical analysis of tissues~\cite{revue-morphometrie, 
morphometrie-comparaison-de-methodes, 
revue-analyse-image-histopathologique, lames, openaccess-1}. A growing 
number of works report the pertinence of such images for diagnosis and 
classification of diseases, e.g.\ tumours~\cite{graphes-gliomes-1, 
graphes-gliomes-2, morphometrie-exemple-colorectal, 
morphometrie-exemple-reperage-noyaux-dans-glioblastomes, openaccess-5}. 
Databases of clinical 
cases~\cite{base-de-donnees-astrocytomes-pediatriques} will include more 
and more digitized tissue images. This growing use of virtual microscopy 
is accompanied by the development of integrated image analysis systems 
offering both virtual slide scanning and automatic image analysis, which 
makes integration into the daily practice of pathologists easier. See 
Ref.~\cite{revue-systemes-de-lames} for a review of some of these 
systems.

 Modern slide scanners produce high magnification microscopy images of 
excellent quality~\cite{virtual-microscopy}, for instance at the 
so-called ``40x'' magnification. They allow much better visualization 
and analysis than lower magnification images. As an example, Figure~1 
shows two portions of a slide at different magnifications, 10x and 40x. 
The benefit of the high magnification for both diagnosis and automated 
image analysis is clear. For instance, the state of the chromatin inside 
the nucleus and the cell morphology, better seen at high magnification, 
are essential to help the clinician distinguish tumorous and 
non-tumorous cells. An accurate, non-pixelated determination of the 
perimeters of the cell nuclei is needed for morphometry and statistics.

 However, this technique involves the manipulation of huge images (of 
the order of 10 billions of pixels for a full-size slide at 
magnification 40x with a single focus level) for which the approach 
taken by most standard software, loading and decompressing the full 
image into RAM, is impossible (a single slice of a full-size slide needs 
of the order of 30~GiB of RAM). As a result, standard open-source 
software such as ImageJ~\cite{imagej}, ImageMagick~\cite{imagemagick} or 
GraphicsMagick~\cite{graphicsmagick} completely fails or is 
prohibitively slow when used on these images. Of course, commercially 
available software exists~\cite{revue-systemes-de-lames}, but it is 
usually quite expensive, and very often restricted to a single operating 
system. It uses proprietary source code, which is a problem if one wants 
to control or check the algorithms and their parameters when doing image 
analysis for research.

 In addition, many automated microscopes or slide scanners store the 
images which they produce into proprietary or poorly documented file 
formats, and the software provided by vendors is often specific to some 
operating system. This leads to several concerns. First, it makes 
research based on digital pathology technically more difficult. Even 
when a project is led on a single site, one has often to use clusters of 
computers to achieve large-scale studies of many full-size slides from 
several 
patients~\cite{morphometrie-exemple-gliomes-diffus-grande-envergure}. 
Since clusters of computers are typically run by open source software 
such as Linux, pathology images stored in non-standard file formats are 
a problem. Furthermore, research projects are now commonly performed in 
parallel in several sites, not to say in several countries, thanks to 
technology such as Grid~\cite{openaccess-3}, and there is ongoing 
efforts towards the interoperability of information systems used in 
pathology~\cite{openaccess-4, wide}. Second, proprietary formats may 
hinder the development of shared clinical 
databases~\cite{base-de-donnees-astrocytomes-pediatriques} and access of 
the general public to knowledge, whereas the citizen should receive 
benefit of public investments. Finally, they may also raise financial 
concerns and conflicts of interest~\cite{openaccess-2}.

 There have been recent attempts to define open, documented, 
vendor-independent software~\cite{openslide, bioformats}, which partly 
address this problem. However, very large images stored in the NDPI file 
format produced by some slide scanners manufactured by Hamamatsu, such 
as the NanoZoomer, are not yet fully supported by such software. For 
instance, LOCI Bio-Formats~\cite{bioformats} is presently unable to open 
images, one dimension of which is larger than 65k, and does not deal 
properly with NDPI files of more than 4~GiB. OpenSlide~\cite{openslide} 
does not currently support the NDPI format. 
NDPI-Splitter~\cite{ndpi-splitter} needs to be run on Windows and 
depends on a proprietary library.

 To address these problems, we have developed open source tools which 
achieve two main goals: reading and converting images in the NDPI file 
format into standard open formats such as TIFF, and splitting a huge 
image, without decompressing it entirely into RAM, into a \emph{mosaic} 
of much smaller pieces (tiles), each of which can be easily opened or 
processed by standard software. All this is realized with high treatment 
speed on all platforms.

\section*{Implementation}

\subsection*{Overview}

 The main software is implemented in the C programming language as 
separate, command-line driven executables. It is independent of any 
proprietary library. This ensures portability on a large number of 
platforms (we have tested several versions of Mac~OS~X, Linux and 
Windows), modularity and ease of integration into scripts or other 
software projects.

 It is complemented by a set of plugins for the public domain software 
ImageJ~\cite{imagej}, implemented in Java, which call the main 
executables in an automatic way to enable an interactive use.

 The LargeTIFFTools and NDPITools are based on the open source 
TIFF~\cite{tiff} and JPEG~\cite{jpeg} or 
libjpeg-turbo~\cite{libjpeg-turbo} libraries. The NDPITools plugins for 
ImageJ are based on the Java API of ImageJ~\cite{imagej, 
imagej-nature-methods} and on the open source software 
Image-IO~\cite{imageio}, and use the Java Advanced Imaging 1.1.3 
library~\cite{jai}.

\subsection*{Basic functions}

 The basic functions are the following. They can be performed even on a 
computer with a modest amount of RAM (see below the ``performance'' 
discussion).

\begin{enumerate}
\item splitting a tiled TIFF file into multiple TIFF files, one for each 
of the tiles (\texttt{tiffsplittiles} program);

\item extracting (``cropping'') quickly a given rectangle of a 
supposedly tiled TIFF file into a TIFF or JPEG file 
(\texttt{tifffastcrop} program);

\item splitting one or several TIFF file(s), possibly very large, into 
mosaic(s), without fully decompressing them in memory 
(\texttt{tiffmakemosaic} program);

\item converting a NDPI file into a standard multiple-image TIFF file, 
tiled if necessary, using upon request the BigTIFF format introduced in 
version~4.0.0 of the TIFF library~\cite{tiff, 
bigtiff-specification-libtiff, bigtiff-specification-awaresystems}, and 
encoding magnification and focus levels as TIFF ``image description'' 
fields (\texttt{ndpi2tiff} program);

\item creating a standard TIFF file for all or part of the magnification 
levels and focus levels present in the given NDPI file (the user can ask 
for specific magnification and focus levels and for a specific 
rectangular region of the image), and, upon request, creating a mosaic 
for each image which doesn't fit into RAM or for all images 
(\texttt{ndpisplit} program). The names of the created files are built 
on the name of the source file and incorporate the magnification and 
focus levels (and, in the case of mosaic pieces, the coordinates inside 
the mosaic).

\end{enumerate}

\subsection*{Mosaics}

 A mosaic is a set of TIFF or JPEG files (the \emph{pieces}) which would 
reproduce the original image if reassembled together, but of manageable 
size by standard software. The user can either specify the maximum 
amount of RAM which a mosaic piece should need to be uncompressed 
(default: 1024~MiB), or directly specify the size of each piece. In the 
first case, the size of each piece is determined by the software. A 
given amount of overlap between mosaic pieces can be requested, either 
in pixels or as a percentage of the image size. This is useful e.g.\ for 
cell counting, not to miss cells which lie on the limit between two 
adjacent pieces.

\subsection*{Usage}

\subsubsection*{Standalone}

 Our tools can be used through the command line (POSIX-like shell or 
Windows command interpreter), and therefore can be very easily 
integrated into scripts or other programs. Depending on the tool, the 
paths and file names of one or several files, in NDPI or TIFF format, 
have to be provided. Options can be added with their arguments on the 
command line to modify the behavior of the programs from its default. 
They are explained in the messages printed by the programs run without 
arguments, in Unix-style man pages, and on the web pages of the project 
(see below in the \emph{Availability and requirements} Section).

 Under the Windows OS, one can click-and-drag the NDPI file icon onto 
the icon of \texttt{ndpi2tiff} or \texttt{ndpisplit}. We provide 
precompiled binaries where frequently-used options are turned on by 
default: e.g.\ \texttt{ndpisplit-mJ.exe} produces a mosaic in JPEG 
format as with option \texttt{-mJ}. The conversion result or mosaic can 
be found in the same directory as the original NDPI image.

\subsubsection*{ImageJ integration}

 In addition to command line use, the \texttt{ndpisplit} program can be 
driven through the \texttt{NDPITools} plugins in ImageJ with a 
point-and-click interface, so that previewing the content of a NDPI file 
at low resolution, selecting a portion, extracting it at high resolution 
and finally opening it in ImageJ to apply further treatments can be done 
in an easy and graphical way. Figure~2 shows a screen shot of 
ImageJ~1.47m after extraction of a rectangular zone from a NDPI file. 
Figure~3 explains what happens when the NDPI file contains several 
levels of focalization: the preview image is displayed as a stack.

 When producing a mosaic, the user can request that pieces be JPEG 
files. Since the \texttt{File > Open} command of versions~1.x of ImageJ 
is unable to open TIFF files with JPEG compression (one has to use 
plugins), this is way to produce mosaics which can be opened by 
click-and-drag onto the window or icon of ImageJ while still saving disk 
space thanks to efficient compression. Figure~4 shows how the mosaic 
production options can be set inside ImageJ through the NDPITools 
plugins.

\section*{Results and Discussion}

  \subsection*{Performance}

 We compare the performance of our tools on several fundamental tasks to 
standard, broadly available software in representative examples and on 
broadly available computers.

\subsubsection*{Making a mosaic from a huge image.}


 We chose an 8-bit RGB colour JPEG-compressed TIFF file of 
103168$\times$63232 pixels originating in the digitization of a 
pathology slide. The original file weighted 975.01~MiB. Loading this 
image entirely into RAM would need at least $3 \times 103168 \times 
63232 = 18.2$~GiB and is presently intractable on most if not all 
desktop and laptop computers of reasonable cost.

 The task was to produce, from this image, a mosaic of 64 pieces so that 
each one needs less than 512~MiB RAM to open.

 On a 3.2~GHz Intel Core~i3 IMac computer with 16~GB of RAM, the 
\texttt{convert} command from ImageMagick (version 6.8.0-7 with quantum 
size 8 bits) was unable to complete the request. GraphicsMagick 
(\texttt{gm convert -crop}; version 1.3.17 with quantum size 8 bits) 
completed the request in 70~min, using 25~GiB of disk space. 
\texttt{tiffmakemosaic} from our LargeTIFFTools completed the request in 
2.5~min.

 To ascertain that this task can be equally achieved even on computers 
with a modest RAM amount, we performed the same task on a 6-year-old 
2.66 GHz Core2Duo Intel IMac with 2~GiB RAM. The task was completed in 
9.0~min.

\subsubsection*{Converting NDPI into TIFF.}


\noindent \emph{Splitting a NDPI file into TIFF files.} A pathology 
sample (6.7~cm$^2$ of tissue) was scanned at magnification 40x and with 
11 focus levels (every 2~microns) by a NanoZoomer, resulting in a 
6.5~GiB file in proprietary NDPI format (called file \texttt{a.ndpi} 
hereafter). On a 2.6~GHz Intel Core i7 Mac~Mini computer with 16~GiB 
RAM, \texttt{ndpisplit} extracted all 55~images (11~focus levels and 
5~magnifications) as independent, single-image TIFF files with JPEG 
compression in 7.11~min. The size of the largest images was $180224 
\times 70144$. The speed was limited only by the rate of I/O transfers 
since the CPU usage of this task was 1.38~min, out of which the system 
used 1.30~min. Executing again the same task straight after the first 
execution took only 0.57~min because the NDPI file was still in the 
cache of the operating system.

 To ascertain that this task can be equally achieved even on computers 
with a modest RAM amount, we made a try on a 6-year-old 2.66 GHz 
Core2Duo Intel PC with 2~GiB RAM running 32-bits Windows XP Pro SP3. The 
original file shown in Figure~1, called \texttt{b.ndpi}, and weighting 
2.07~GiB (largest image: $103168 \times 63232$ pixels), was split into 
independent TIFF files in 2.2~min without swapping.

 In comparison, the LOCI Bio-Formats plugins for 
ImageJ~\cite{bioformats}, in its version 4.4.6 with ImageJ~1.43m, was 
not able to open the images in file~\texttt{a.ndpi} even at low 
resolution. \pb

\noindent \emph{Converting a NDPI file into a multiple-images TIFF 
file.} Alternatively, the same proprietary-format file \texttt{a.ndpi} 
was converted into a multiple-images TIFF file with \texttt{ndpi2tiff}. 
On the same computer as before, the conversion time was 7.0~min. Here 
again, the speed of the process is limited only by the rate of I/O 
transfers since the conversion took only 30~s if performed when the NDPI 
file was still in the cache of the operating system.

 Since the resulting TIFF file could not store all 55~images in less 
than 4~GiB, we passed the option~\texttt{-8} on the command line to 
\texttt{ndpi2tiff} to request using the BigTIFF format extension. The 
specifications of this extension to the TIFF standard, discussed and 
published before 2008~\cite{bigtiff-specification-libtiff, 
bigtiff-specification-awaresystems}, are supported by LibTIFF as of 
version~4.0.0~\cite{tiff}, and therefore by the abundant image viewing 
and manipulation software which relies on LibTIFF. If the use of the 
BigTIFF format extension would have impeded the further exploitation of 
the produced TIFF file, we could have simply used \texttt{ndpisplit} as 
above. Or we could have called the \texttt{ndpi2tiff} command several 
times, each time requesting extraction of a subset of all images by 
specifying image numbers after the file name, separated with commas, as 
in \texttt{a.ndpi,0,1,2,3,4}.

\subsubsection*{Extracting a small region from a huge image.}

This task can be useful to visualize at full resolution a region of 
interest which the user has selected on a low-magnification preview 
image. Therefore, it should be performed as quickly as possible. \pb

\noindent \emph{From a TIFF file.}

\noindent The task was to extract a rectangular region of size 
$256\times256$ pixels situated at the bottom right corner of huge TIFF 
images and to save it as an independent file. The source images were 
single-image TIFF files using JPEG compression. Table~1 compares the 
time needed to complete the task with \texttt{tifffastcrop} from our 
LargeTIFFTools and with several software tools, on increasingly large 
TIFF files. Tests were performed on a 2.6~GHz Intel Core~i7 Mac Mini 
computer with 16~GB of RAM and used GraphicsMagick~1.3.17, 
ImageMagick~6.8.0-7 and the utility \texttt{tiffcrop} from 
LibTIFF~4.0.3. Noticeably, when treating the largest image, 
GraphicsMagick needs 50~GiB of free disk space, whereas 
\texttt{tifffastcrop} doesn't need it. \pb

\noindent \emph{From a NDPI file.}

\noindent The task was to extract a rectangular region of size 
$256\times256$ from one of the largest images of the file 
\texttt{a.ndpi} (size $180224 \times 70144$). On a 2.6~GHz Intel Core~i7 
Mac Mini computer with 16~GB of RAM, the execution time was 0.12~s for 
one extract, and in average 0.06~s per extract in a series of 
20~extracts with locations drawn uniformly at random inside the whole 
image.

\subsection*{Applications}

\subsubsection*{Integration in digital pathology image servers or 
virtual slide systems}

 The NDPITools are being used in several other software projects:

\begin{itemize}
\item in a system for automatic blur detection~\cite{stack-or-trash, 
these-david-ameisen}.

\item in WIDE~\cite{wide}, to deal with NDPI files. WIDE is an 
open-source biological and digital pathology image archiving and 
visualization system, which allows the remote user to see images stored 
in a remote library in a browser. In particular, thanks to the feature 
of high-speed extraction of a rectangular region by \texttt{ndpisplit}, 
WIDE saves costly disk space since it doesn't need to store TIFF files 
converted from NDPI files in addition to the latter.
\end{itemize}

\subsubsection*{Exploiting a large set of digital slides}

 In the framework of a study about invasive low-grade oligodendrogliomas 
reported elsewhere~\cite{lames}, we had to deal with 303 NDPI files, 
occupying 122~GiB. On a 3.2~GHz Intel Core i3 IMac computer with 16~GB 
RAM, we used \texttt{ndpisplit} in a batch work to convert them into 
standard TIFF files, which took only a few hours. The experimental 
\texttt{-s} option of \texttt{ndpisplit} was used to remove the blank 
filling between scanned regions, resulting in an important disk space 
saving and in smaller TIFF files (one for each scanned region) which 
where easier to manipulate afterwards. Then, for each sample, 
Preview.app and ImageJ were used to inspect the resulting images and 
manually select the regions of clinical interest. The corresponding 
extracts of the high magnification images were the subject of automated 
cell counting and other quantitative analyses using ImageJ. In 
particular, we collected quantitative data about edema or tissue 
hyperhydration~\cite{lames}. This quantity needed a specific image 
analysis procedure which is not offered by standard morphometry software 
and, unlike cell density estimates, could not be retrieved by sampling a 
few fields of view in the microscope. Therefore, virtual microscopy and 
our tools were essential in this study.

\subsubsection*{Study of a whole slide of brain tissue invaded by an 
oligodendroglioma}

 To demonstrate the possibility to do research on huge images even with 
a modest computer, we chose a 3-year-old MacBook Pro laptop computer 
with 2.66~GHz Intel Core 2 Duo and 4~GiB of RAM. We used ImageJ and the 
NDPITools to perform statistics on the upper piece of tissue on the 
slide shown in Figure~1.

 Since the digital slide \texttt{b.ndpi} weighted 2.07~GiB, with a high 
resolution image of $103168 \times 63232$~pixels, it was not possible to 
do the study in a straightforward way. We opened the file 
\texttt{b.ndpi} as a preview image with the command \texttt{Plugins > 
NDPITools > Preview NDPI...} and selected on it the left tissue sample. 
Then we used the command \texttt{Plugins > NDPITools > Custom extract to 
TIFF / Mosaic...} and asked for extraction as a mosaic of 16 JPEG files, 
each one needing less than 1 GiB of RAM to open, and with an overlap of 
60 pixels. This was completed within a few minutes. Then we applied an 
ImageJ macro to each of the 16 pieces to identify the dark cell nuclei 
(those with high chromatin content), based on thresholding the 
luminosity values of the pixels, as shown in Figure~1. It produced text 
files with the coordinates and size of each cell nucleus.

 Out of the 154240 identified nuclei, 1951 were positioned on the 
overlapping regions between pieces. Using the overlap feature of our 
tools enabled to properly detect these nuclei, since they would have 
been cut by the boundary of the pieces of the mosaic in absence of 
overlap. We avoided double counting by identifying the pairs of nuclei 
situated in the overlapping regions and which were separated by a 
distance smaller than their radius.

 As shown in earlier studies~\cite{graphes-gliomes-1, graphes-gliomes-2, 
revue-analyse-image-histopathologique}, these data can be used for 
research and diagnosis purposes. As an example, Figure~5 shows the 
distribution of the distance of each cell nucleus to its nearest 
neighbor. Thanks to the very high number of analyzed cell nuclei, this 
distribution is obtained with an excellent precision.

\section*{Conclusions}

 The LargeTIFFTools, NDPITools and NDPITools plugins for ImageJ achieve 
efficiently some fundamental functions on large images and in particular 
digital slides, for which standard open source software fails or 
performs badly. They enable both the clinician to examine a single slide 
and the bioinformatics research team to perform large-scale analysis of 
many slides, possibly on computer 
grids~\cite{morphometrie-exemple-gliomes-diffus-grande-envergure}.

 To date, the LargeTIFFTools have been downloaded from more than 388 
different IP addresses, the NDPITools from more than 1361 addresses, and 
the ImageJ plugins from more than 235 addresses. Table~2 lists the 
distribution of the target platforms among the downloads of the binary 
files. It shows a broad usage of the different platforms by the 
community, emphasizing the importance of cross-platform, open source 
tools.

 We have explained how the software was used to study some microscopic 
properties of brain tissue when invaded by an oligodendroglioma, and we 
have given an illustrative application to the analysis of a whole-size 
pathology slide. This suggests other promising applications.

\section*{Availability and requirements}

\noindent a. LargeTIFFTools

\begin{itemize}
\item \textbf{Project name:} LargeTIFFTools
\item \textbf{Project home page:} 
\url{http://www.imnc.in2p3.fr/pagesperso/deroulers/software/largetifftools/}
\item \textbf{Operating system(s):} Platform independent
\item \textbf{Programming language:} C
\item \textbf{Other requirements:} libjpeg, libtiff
\item \textbf{License:} GNU GPLv3
\end{itemize}

\noindent b. NDPITools

\begin{itemize}
\item \textbf{Project name:} NDPITools
\item \textbf{Project home page:} 
\url{http://www.imnc.in2p3.fr/pagesperso/deroulers/software/ndpitools/}
\item \textbf{Operating system(s):} Platform independent
\item \textbf{Programming language:} C
\item \textbf{Other requirements:} ---
\item \textbf{License:} GNU GPLv3
\end{itemize}

 For the convenience of users, precompiled binaries are provided for 
Windows (32 and 64 bits), Mac~OS~X and Linux. \pb

\noindent c. NDPITools plugins for ImageJ

\begin{itemize}
\item \textbf{Project name:} NDPITools plugins for ImageJ
\item \textbf{Project home page:} 
\url{http://www.imnc.in2p3.fr/pagesperso/deroulers/software/ndpitools/}
\item \textbf{Operating system(s):} Platform independent
\item \textbf{Programming language:} Java
\item \textbf{Other requirements:} ImageJ 1.31s or higher, Ant, JAI 1.1.3
\item \textbf{License:} GNU GPLv3
\end{itemize}

  \ifthenelse{\boolean{publ}}{\small}{}

\section*{Competing interests}

The authors declare that they have no competing interests.
    
\section*{Authors contributions}

CD wrote the paper. ML conceived and implemented a first version of the 
integration into ImageJ as a \emph{toolset} of macros. CD implemented 
the software and wrote the documentation. CG, AG and ML contributed 
suggestions to the software. CD, DA, AG and ML performed software tests. 
CD, MB, CG, AG and ML selected and provided histological samples. CD 
performed the statistical analysis of the sample slide. All authors 
reviewed the manuscript. All authors read and approved the final 
manuscript.

\section*{Acknowledgements}

We thank F.~Bouhidel and P.~Bertheau for their help with the slide 
scanner of the Pathology Laboratory of the Saint-Louis Hospital in 
Paris, and C.~Klein (Imaging facility, Cordeliers Research Center -- 
INSERM U872, Paris) for tests and suggestions.

The computer, CPU, operating system, and programming language 
names quoted in this article are trademarks of their respective owners.
 

{\ifthenelse{\boolean{publ}}{\footnotesize}{\small}
 \bibliographystyle{bmc_article}  
  \bibliography{ndpitools} }     


\ifthenelse{\boolean{publ}}{\end{multicols}}{}



\section*{Figures}

\begin{minipage}{\linewidth}
\begin{center}
\includegraphics[width=\linewidth]{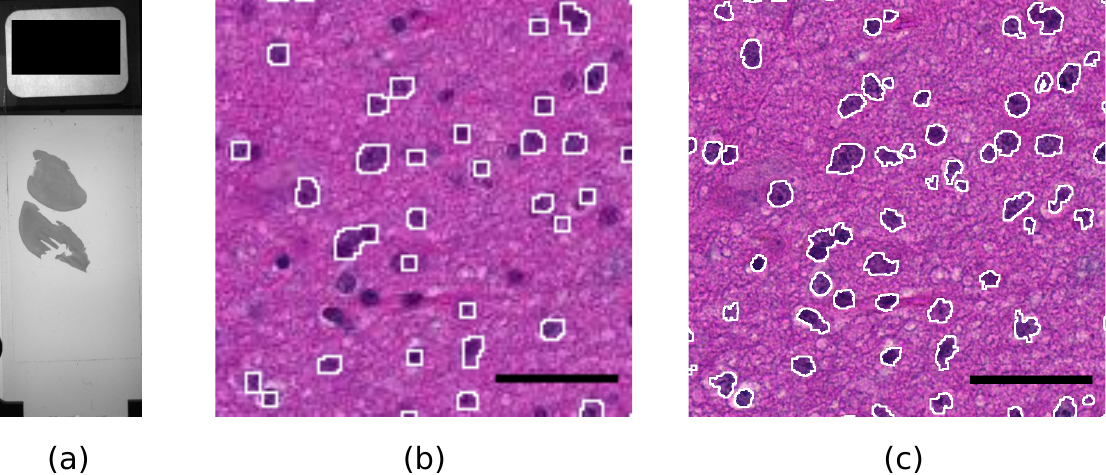}
\end{center}
  \subsection*{Figure 1 - A sample slide.}
(a): macroscopic view of the whole slide (the black rectangle on the 
left is 1x2~cm). (b,c): Influence of the magnification on the quality of 
results. (b): a portion of the slide scanned at magnification level 10x. 
The white contours show the result of an automatic detection of the dark 
cell nuclei with the ImageJ software. A significant fraction of the cell 
nuclei is missed and the contours are rather pixelated. (c): the same 
portion of the slide scanned at magnification 40x. The white contours 
show the result of the same automatic detection. Almost all cell nuclei 
are detected and the shapes of the contours are much more precise. Scale 
bar: 4~$\mu$m.
\end{minipage}

\mbox{}\bigskip\mbox{}

\noindent \begin{minipage}{\linewidth}
\begin{center}
\includegraphics[width=\linewidth]{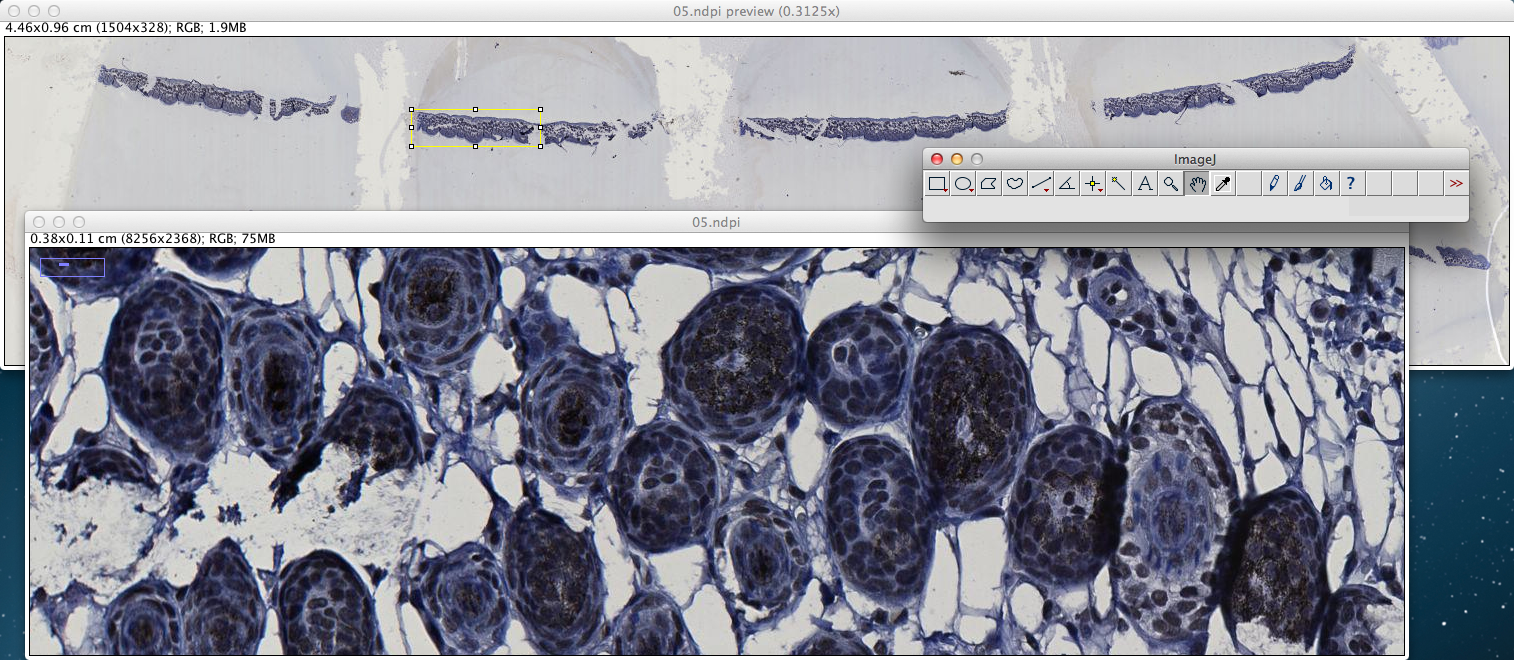}
\end{center}
  \subsection*{Figure 2 - A typical session using ImageJ and the 
NDPITools plugins.}
 A NDPI file has been opened with the NDPITools plugins and it is 
displayed as a preview image (image at largest resolution which still 
fits into the computer's screen) --- top window. A rectangular region 
has been selected and extracted as a TIFF image, then opened --- bottom 
window.
\end{minipage}

\noindent \begin{minipage}{\linewidth}
\begin{center}
\includegraphics[width=0.45\linewidth]{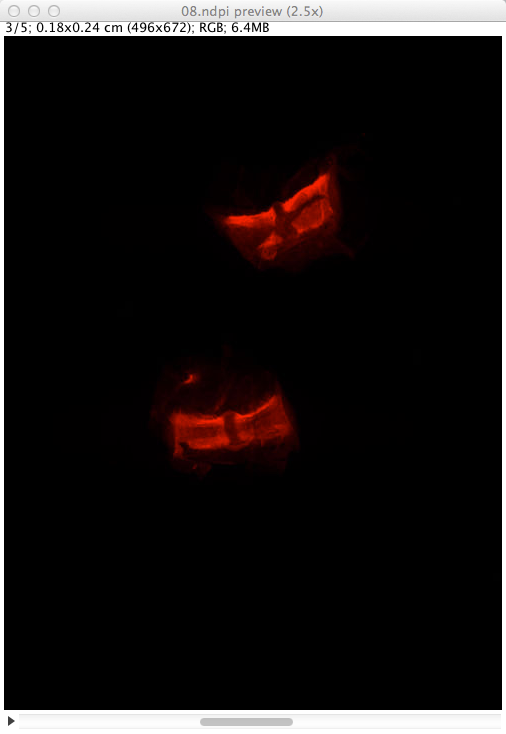}
\end{center}
  \subsection*{Figure 3 - Preview image of a NDPI file with several 
focalization levels in ImageJ.}
 The NDPI file \texttt{08.ndpi} contains images at 5 different 
focalization levels. Therefore, its preview image is displayed as a 
stack of 5 images.
\end{minipage}

\noindent \begin{minipage}{\linewidth}
\begin{center}
\includegraphics[width=0.45\linewidth]{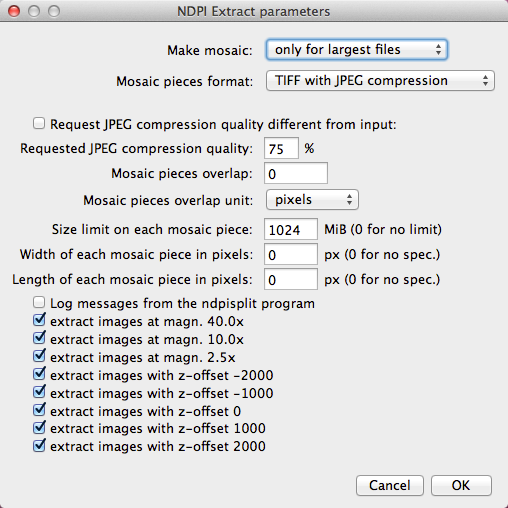}
\end{center}
  \subsection*{Figure 4 - Dialog box for customized extraction in ImageJ 
from an NDPI file with production of a mosaic.}
 The dialog box shows some options which can be customized while 
producing a mosaic from a rectangular selection of a NDPI file preview 
image (here, using the file previewed in Figure~3).
\end{minipage}

\mbox{}\bigskip\mbox{}

\noindent \begin{minipage}{\linewidth}
\begin{center}
\includegraphics[width=0.5\linewidth]{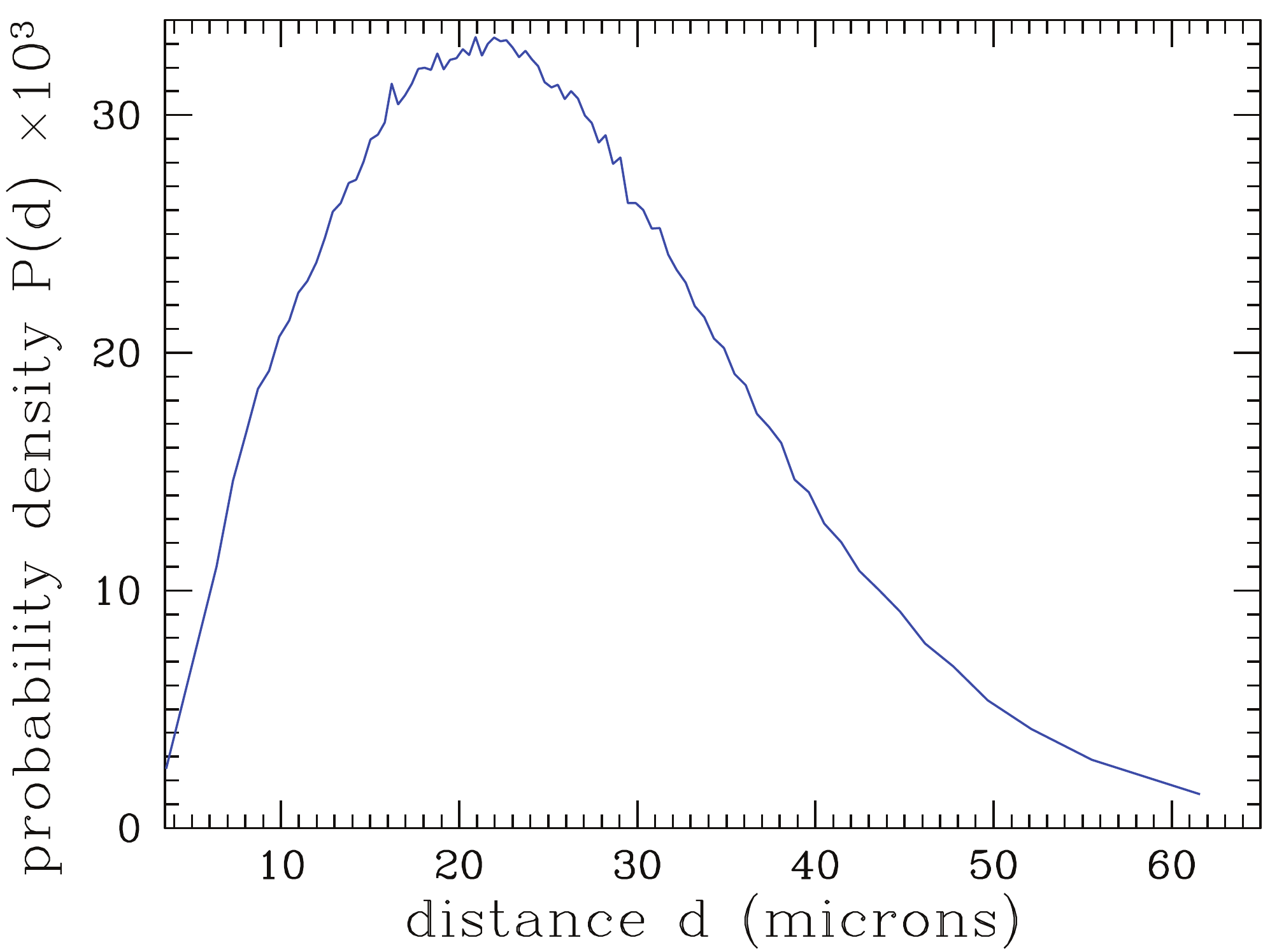}
\end{center}
  \subsection*{Figure 5 - Statistical properties of the cell nuclei with 
high chromatin content in the tissue sample of Figure~1.} The positions 
of the 154240 identified nuclei were obtained from the analysis with 
ImageJ of the digital slide on a laptop computer. Since the slide was 
too large to fit into the computer's memory, it was turned into a mosaic 
of 16 pieces with overlap of 60 pixels, and each piece underwent 
automated analysis independently. Then the results were aggregated. The 
graph shows the probability density function of the distance of a cell
nucleus to its nearest neighbor in the whole sample.
\end{minipage}


\section*{Tables}
  \subsection*{Table 1 - Speed comparison of software to extract a 
$256\times256$ rectangle from a huge TIFF image}
 Time needed (or indication of failure when the task was not completed) 
by several software tools to extract a rectangular region of size 
$256\times256$ pixels situated at the bottom right corner of huge TIFF 
images, and to save it as an independent file. The input images were 
single-image tiled TIFF files using JPEG compression. Their dimensions 
are indicated in the top row. The computer used was a 2.6~GHz Intel 
Core~i7 Mac Mini with 16~GiB of RAM and more than 100~GiB of free hard 
disk. The tested software tools were, from top to bottom, 
\texttt{tifffastcrop} from our LargeTIFFTools, GraphicsMagick~1.3.17, 
ImageMagick~6.8.0-7 and the utility \texttt{tiffcrop} from 
LibTIFF~4.0.3. \par \mbox{}
    \par
    \mbox{
\begin{tabular}{|c||c|c|c|}
\hline
Image size (px) & \small{$11264\times4384$} & \small{$45056\times17536$} & 
\small{$180224\times70144$} \\
\hline \hline
\texttt{tifffastcrop} & 0.30~s & 0.30~s & 0.30~s  \\
\hline
GraphicsMagick & 0.74~s & 23.6~s & $> 80$~min \\
\hline
ImageMagick & 1.18~s & 236~s & failed  \\
\hline
\texttt{tiffcrop} & 0.50~s & failed & seg. fault  \\
\hline
\end{tabular}
      }

  \subsection*{Table 2 - Downloads of the NDPITools}
    Distribution of the downloads (unique IP address) of the precompiled 
binaries of the NDPITools between March 2012 and April 2013 \par \mbox{}
    \par
    \mbox{
\begin{tabular}{|c|c|c|c|}
\hline
Windows (32 bits) & Windows (64 bits) & Linux & Mac OS X\\
\hline
483 & 542 & 217 & 285 \\
\hline
\end{tabular}
      }

\end{bmcformat}
\end{document}